\documentclass{desyproc}

\begin{document}
\title{Search for a new short-range spin-dependent force with polarized Helium 3}

\author{{\slshape Mathieu Guigue$^1$, David Jullien$^{2}$, Alexander K. Petukhov$^2$, Guillaume Pignol$^1$}\\[1ex]
$^1$LPSC, Universit\'e Grenoble-Alpes, CNRS/IN2P3, Grenoble, France\\
$^2$Institut Laue Langevin, 6 Rue Jules Horowitz, 38000 Grenoble, France}

\contribID{guigue\_mathieu}

\desyproc{DESY-PROC-2014-XX}
\acronym{Patras 2014} 
\doi  

\maketitle

\begin{abstract}
Measuring the depolarization rate of a $^3$He hyperpolarized gas is a sensitive method to probe hypothetical short-range spin-dependent forces. 
A dedicated experiment is being set up at the Institute Laue Langevin in Grenoble to improve the sensitivity. 
We presented the status of the experiment at the 10th PATRAS Workshop on Axions, WIMPs and WISPs.
\end{abstract}

\section{Probing short-range spin-dependent interaction}

Numerous theories beyond the Standard Model of particle physics predict the existence of new light scalar bosons such as the Axions theory \cite{Moody1984} developed to solve the strong CP problem.
The exchange of a new scalar boson of mass $m _{\phi}$  between a polarized probe particle and an unpolarized source particle, separated by $r$,  would mediate a short-range monopole-dipole interaction defined by the potential:
\begin{equation}\label{eq:dipmonointeraction}
 V=g_s^N g_p ^N \frac{\hbar \widehat{\sigma}.\widehat{r}}{8\pi Mc}\left(\frac{m_{\phi}}{r}+\frac{1}{r^2}\right)\exp (-m_{\phi} r)
\end{equation}
where $\hbar\widehat{\sigma}/2$ is the spin of the probe particle, $M$ the mass of the polarized particle, $g_s^N$ and $ g_p^N$ the coupling constant at the vertices of polarized and unpolarized particles corresponding to a scalar and a pseudoscalar interactions. 
Finding a new boson or a new short-range interaction interaction would be a important discovery in fundamental physics, since it could solve problems such as the nature of the Dark Matter (as a WISP candidate \cite{Arias2012}). 
This new force is thus actively searched for around the world through different kind of experiments (using torsion-balance, studying the Newton's inverse square law or looking at bouncing ultracold neutrons \cite{Antoniadis2011}).

Consider a spherical cell filled with polarized $^3$He atoms with a gyromagnetic ratio $\gamma$ immersed in a static magnetic field $B_0$. 
In the case of a light boson with $m_{\phi} \lesssim 1\, \rm{eV}$, the axionlike interaction given by Eq. (\ref{eq:dipmonointeraction}) acts like a macroscopic pseudomagnetic field of typical size $\lambda = \frac{\hbar}{m_{\phi}c}\gtrsim 1\, \rm{\mu m}$, which is generated by the glass walls of the cell:
\begin{eqnarray}\label{defn:magField-inhom}
b_{\rm{NF}}(x)&=&\frac{\hbar\lambda}{2\gamma m_n}Ng_sg_p \left( 1-e^{-d/\lambda}\right) e^{-x/{\lambda}},
\end{eqnarray}
where $N$ is the nucleon density, $m_n$ the nucleon mass, $x$ the distance to the cell wall and $d$ the thickness of the  wall. 
This new pseudomagnetic field acting close and perpendicular to the surface of the cell would provoke the depolarization of the gas, whose rate is, for a pressure of several bars and typical cell sizes $L$ much larger than $\lambda$:
\begin{equation}\label{GammaNF}
\Gamma _{1\rm{,NF}} \approx \sqrt{\frac{2}{\gamma B_0}}\frac{\lambda ^2 \gamma ^2 b_a^2}{L},
\end{equation}
where $b_a$ is the the prefactor of the exponential in Eq. \eqref{defn:magField-inhom}.
This new relaxation channel adds to the the natural depolarization $\Gamma _1$ of the gas, which results from the three main contributions: the relaxation rate induced by collisions of polarized atoms on the cell walls  $\Gamma _w$, the relaxation rate due to interparticles collisions $\Gamma _{dd}$ and the depolarization rate due to the motion of polarized particles in an inhomogeneous magnetic field $\Gamma _m$.
While $\Gamma _1$ behaves as $a + \frac{b}{B_0}+\frac{c}{B_0^2}$, the relaxation rate $\Gamma _{1,\rm{NF}}$ is proportional to $1/\sqrt{B_0}$ which is a non-standard dependence on the holding magnetic field $B_0$.
The existence of a new short-range spin-dependent interaction can then be  extracted from measurements of the relaxation rate at different values of the holding magnetic field.

A first test experiment \cite{Antoniadis2011,Petukhov2010} measuring the spin longitudinal depolarization rate $\Gamma_1$ as a function of the applied field $B_0$  was performed in 2010 to demonstrate the sensitivity of the method. 
A new dedicated experiment is set up at the Institute Laue-Langevin, improving both (i) the magnetic environment of the experiment and (ii) the measurement of the decay of polarization.

\section{An improvement of the magnetic field environment}

\begin{wrapfigure}{l}{0.65\textwidth}
\centerline{\includegraphics[scale=0.27]{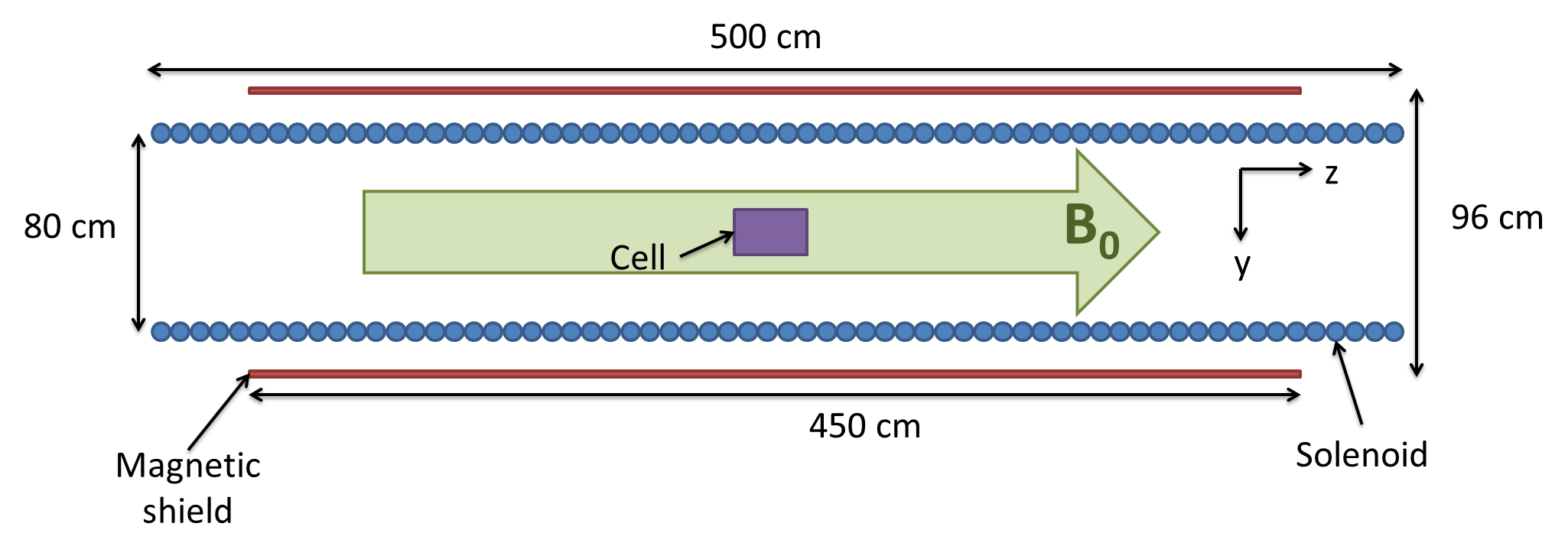}}
\caption{\label{fig:setup} Scheme of the apparatus. }
\end{wrapfigure}
In order to suppress the magnetic field inhomogeneity depolarization channel $\Gamma _m$, the holding magnetic field should be as homogeneous as possible.
The apparatus, represented on Fig. \ref{fig:setup}, is composed of a 5 m long and 80 cm diameter solenoid which provides a very homogeneous magnetic field.
In order to shield its center from external magnetic fields, this solenoid is inserted into a $\mu$-metal magnetic shield of 96 cm diameter and 4 m long from "n-nbar" experiment \cite{Bitter1991a} which measured at the Institut Laue Langevin the neutron-antineutron oscillation.
\begin{figure}[hbtp]
\centerline{
\includegraphics[scale=0.22]{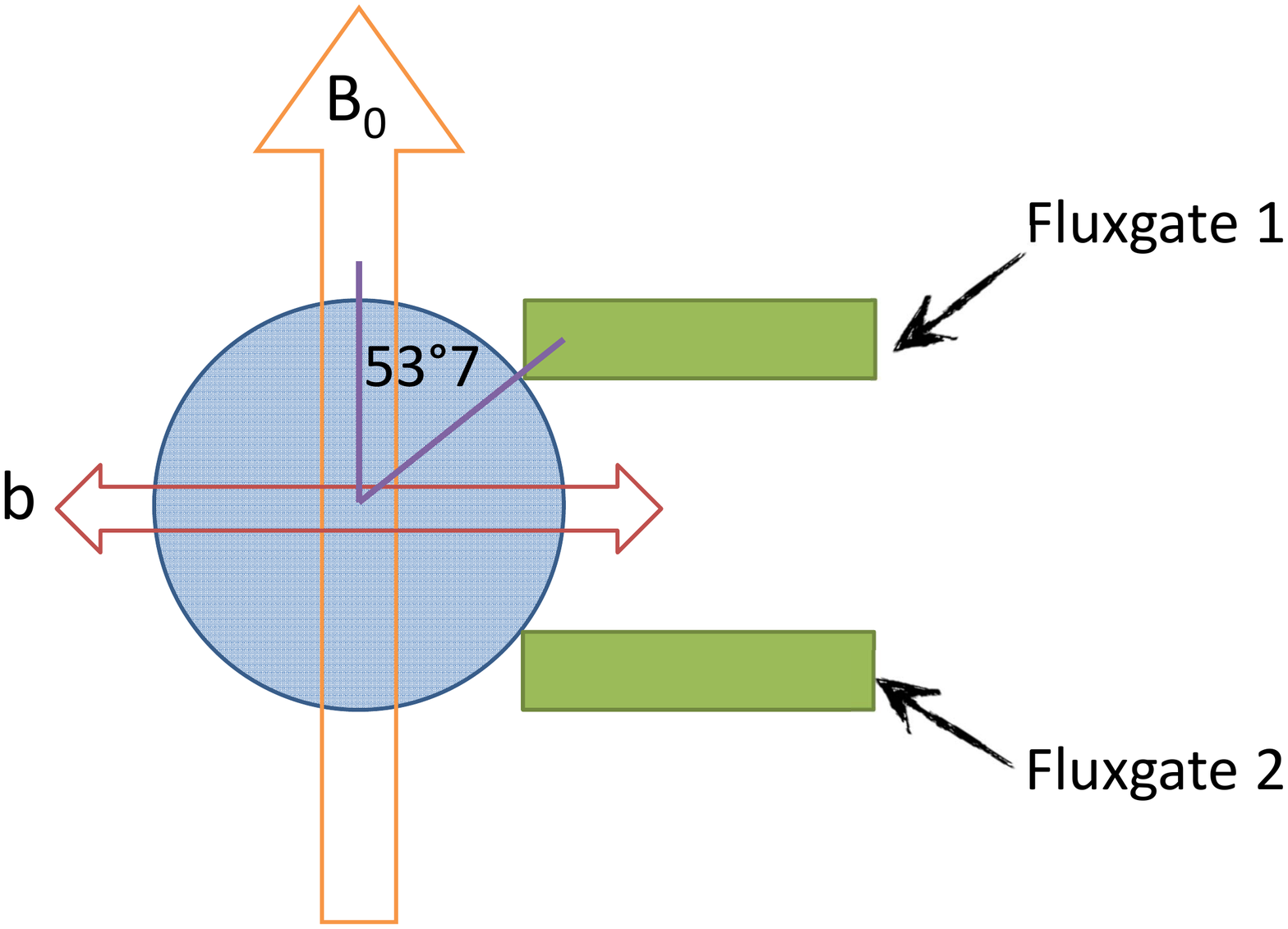}
\includegraphics[scale=0.4]{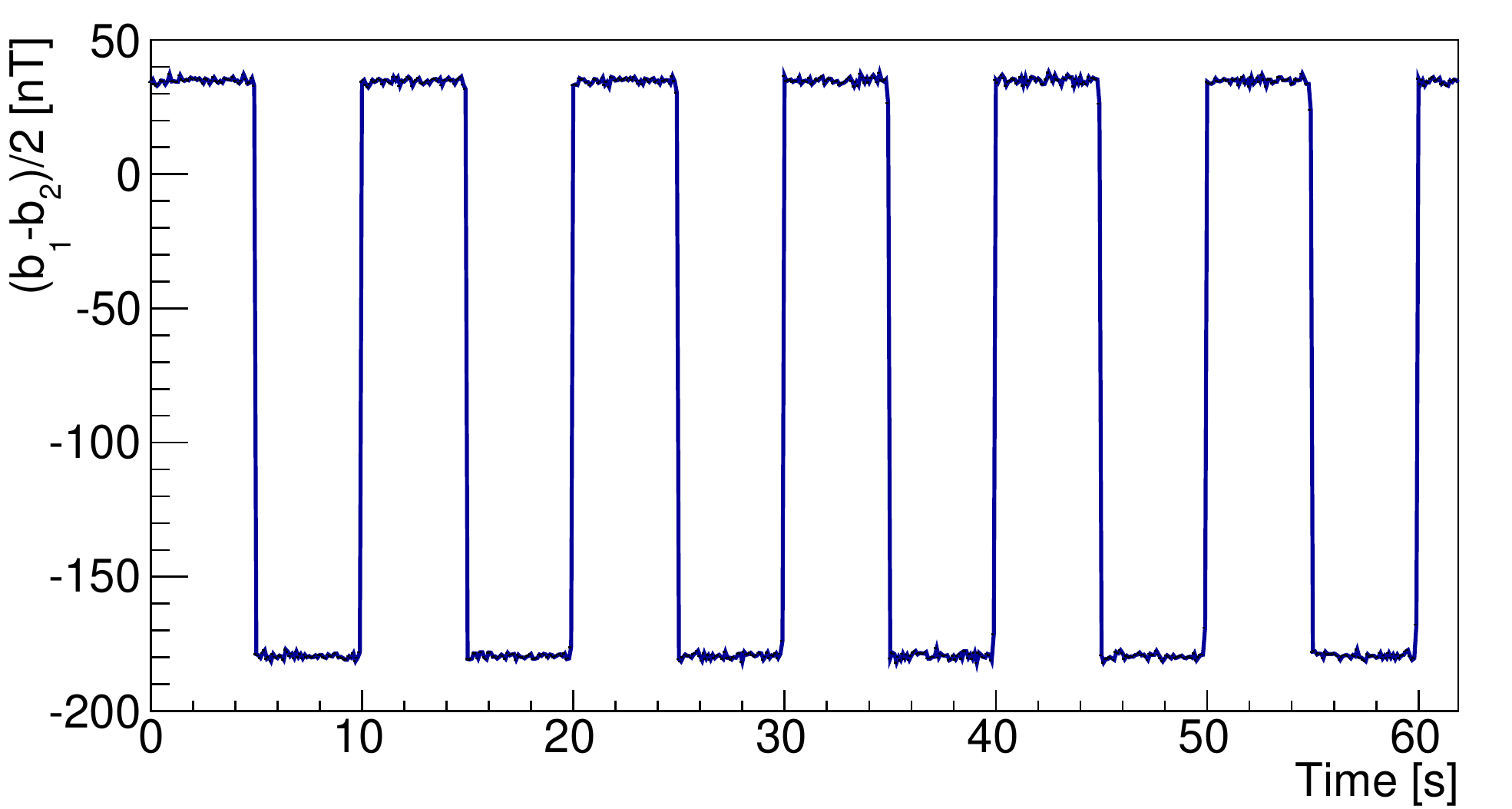}
}
\caption{\label{fig:SF-apparatus} (Left) Scheme of the spin-flipping and measurement apparatus.
The spherical cell is positioned in contact with the two magnetometers inside the spin-flip coil which generates an oscillating magnetic field transversely $b$ to the holding magnetic field $B_0$.
(Right) Typical sequence of measurement of the transverse magnetic field generated by a cylindrical cell at 4 bar with two fluxgate magnetometers.
The upper (lower) points correspond to the spin up (down) state of the gas.}
\end{figure}

In addition, we mapped the inner magnetic field using a three-axis fluxgate magnetometer, in order to extract the transverse gradients $g_\perp$ for different $B_0$ settings.
Typically, $g_{\perp}$ is about $1.5\, \rm{nT/cm}$ to $2\, \rm{nT/cm}$ for magnetic field from $2\, \rm{\mu T}$ to $80\, \rm{\mu T}$.
The magnetic relaxation time $T _m=1/\Gamma _m$ is then expected to be longer than $90\, \rm{h}$.
At low field ($B_0 = 3\, \rm{\mu T}$), this is a factor 100 better than the previous experiment.
This improvement will directly affect the duration of the experiment and so the precision of the relaxation rate measurement.

\section{The polarization measurement technique}

\begin{wrapfigure}{l}{0.52\textwidth}
\centering
\includegraphics[scale=0.38]{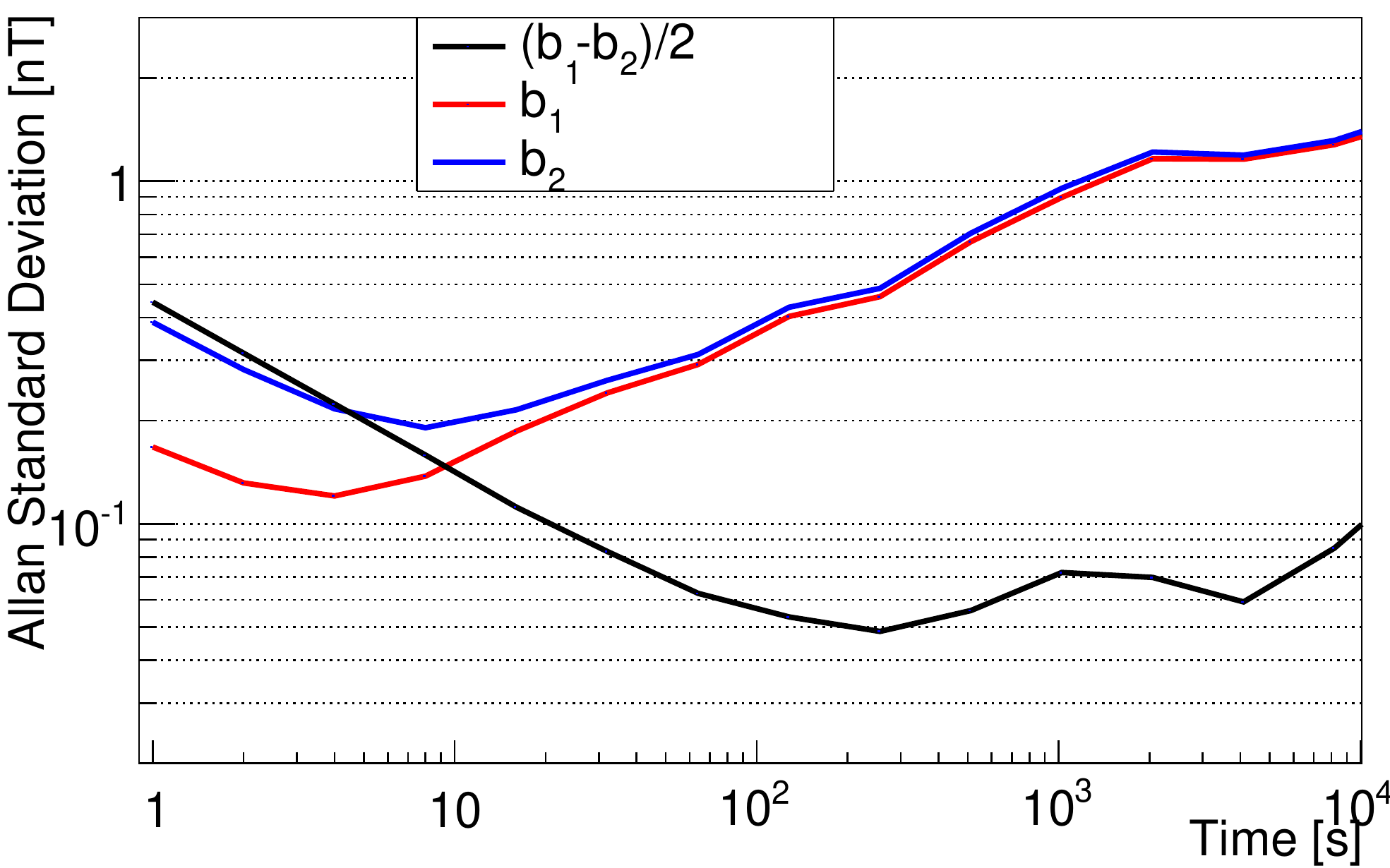}
\caption{\label{ASD} Allan Standard Deviation of the measurements of the transverse magnetic field generated by the solenoid with the two fluxgate magnetometers (in blue and red) and the Allan Standard Deviation of the differential measurement (in black).}
\end{wrapfigure}
Among all the way to measure the polarization and the relaxation rate of an hyperpolarized gas, the nuclear magnetic resonance (NMR) is the most widely used technique. 
One can measure at the percent level the amplitude of the NMR response signal or can also measure the frequency shift, both proportional to the polarization. 

A better precision for the longitudinal relaxation rate measurement can be achieved with a direct polarimetry technique \cite{Wilms1997}: it consists in measuring, with two fluxgate magnetometers, the magnetic field generated by the gas itself which can be of order of tenths of $\rm{nT}$ at pressures of several bars.
The magnetometers sensors are at a place where the dipolar field induced by the polarized gas is transverse relative of the $B_0$ field.

Applying spin-flips with a transverse oscillating magnetic field (Fig. \ref{fig:SF-apparatus}) to reverse the polarization, one can remove magnetic offsets induced by long-term drift of the holding magnetic field or the misalignment of the magnetometers axis with $B_0$.
Fig. \ref{fig:SF-apparatus} shows typical sequences of "up-down-down-up" measurements of the magnetic field induced by a spherical cell at 1 bar with the two magnetometers.

Fig. \ref{ASD} shows the Allan Standard Deviation (ASD) of the holding field measurements with each magnetometers and of differential measurements.
Since the magnetometers measure the same signal with an opposite sign, a  differential acquisition between the two fluxgate magnetometers improves the precision of the measurement by suppressing the long-time correlated fluctuations of the holding magnetic field.
Since the Allan Standard Deviation of the fluxgate magnetometers is $50$ pT at 1 s when they are exposed to no magnetic field, the uncertainty on the polarization measurement mainly comes from the the holding magnetic field instability.
This deviation is about 0.4 nT at 1 s; then, for a 1 bar cell of 70 \% polarized helium, the typical magnetic field generated by the cell is 30 nT.

\begin{wrapfigure}{r}{0.43\textwidth}
\centering
\includegraphics[scale=0.35]{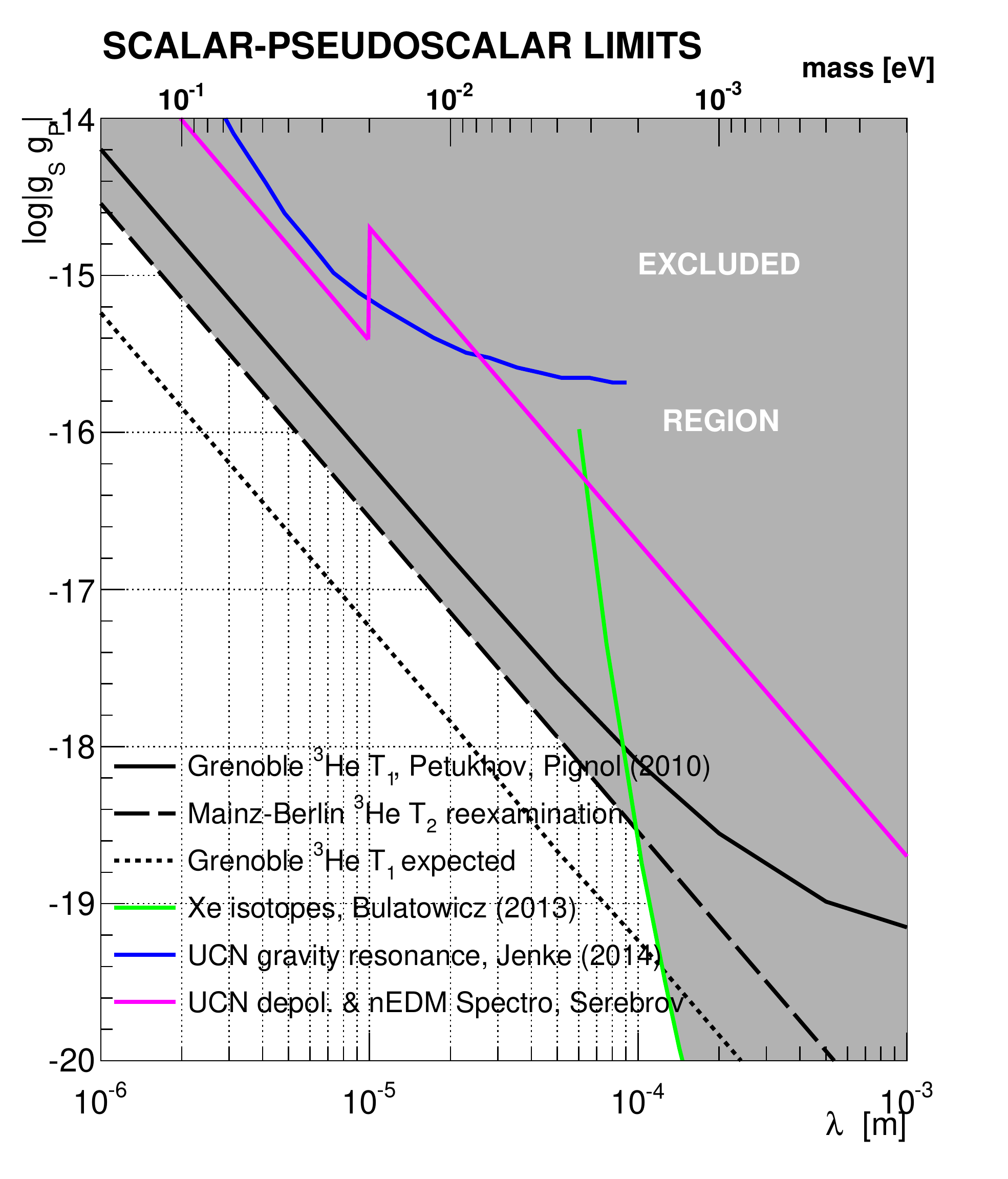}
\caption{\label{exclusion_figure} Constraints on the coupling constant product of axion-like particles to nucleons $g_s^Ng_p^N$ as a function of the range $\lambda$ of the macroscopic interaction.
Bold black line, from $^3$He $T_1$ measurement \cite{Petukhov2010}; long-dashed line, $^3$He $T_2$ reexamination \cite{Petukhov2010}; short-dashed line, expected constraint from upgraded setup (present work); light green, from \cite{Bulatowicz2013}; blue, from \cite{Jenke2014}; pink, from \cite{Serebrov2009,Serebrov2010}.}
\end{wrapfigure}

Since several measurements of magnetic field are performed for a single relative polarization determination, the relative uncertainties on the polarization is about $0.3\, \%$ which leads to relative uncertainties on relaxation rates lower than $1\, \%$.

\section{Expected constraints}

The high quality of the magnetic environment decreases the dominant depolarization contribution at low magnetic field and the direct polarization measurement technique increases our sensitivity to any deviation from the classical behaviour of the longitudinal relaxation rate with the holding magnetic field $B_0$.
From the expected $\Gamma _1$ curve as a function of $B_0$, one can extract the expected sensitivity  (shown on Fig. \ref{exclusion_figure}) of the coupling constants product $g_s^Ng_p^N$ from a new depolarization channel \eqref{GammaNF}.
 

\begin{footnotesize}

\end{footnotesize}


\end{document}